\documentclass[12pt]{iopart}
\usepackage{rotating}
\usepackage{graphicx}
\usepackage{latexsym}
\usepackage{iopams}
\usepackage{bm}

\newcommand{\RP}{\ensuremath{\mathcal{R}}}
\newcommand{\PP}{\ensuremath{\mathcal{P}}}

\begin{document}

\title[Punctured staircase polygons]
{The perimeter generating function of punctured staircase polygons}

\author{Anthony J. Guttmann and Iwan Jensen}
\address{
ARC Centre of Excellence for Mathematics and Statistics of Complex Systems, \\
Department of Mathematics and Statistics, 
The University of Melbourne, Victoria 3010, Australia}

\date{\today}

\ead{tonyg,I.Jensen@ms.unimelb.edu.au} 

\begin{abstract}
Using a simple transfer matrix approach we have derived very long series expansions 
for the perimeter generating function of punctured staircase polygons (staircase polygons
with a single internal staircase hole). We find that all the terms in the generating function
can be reproduced from a linear Fuchsian differential equation of order 8. We perform an 
analysis of the properties of the differential equation. 
\end{abstract}

\submitto{\JPA}


\maketitle

\section{Introduction}

A well-known long standing problem in combinatorics and statistical mechanics is
to find the generating function for self-avoiding polygons (or walks) on a two-dimensional
lattice, enumerated by perimeter. Recently, we have gained a greater understanding 
of the difficulty of this problem, as Rechnitzer \cite{AR03a} has proved that the 
(anisotropic) generating function for square lattice self-avoiding polygons is not 
differentiably finite \cite{RPS80a}, as had been conjectured earlier 
on numerical grounds \cite{Guttmann2001}. 
That is to say, it cannot be expressed as the solution of an ordinary differential equation
with polynomial coefficients. There are many simplifications of this problem that
are solvable \cite{BM96a}, but all the simpler models impose an effective directedness 
or equivalent constraint that reduces the problem, in essence, to a one-dimensional problem.

\begin{figure}
\begin{center}
\includegraphics{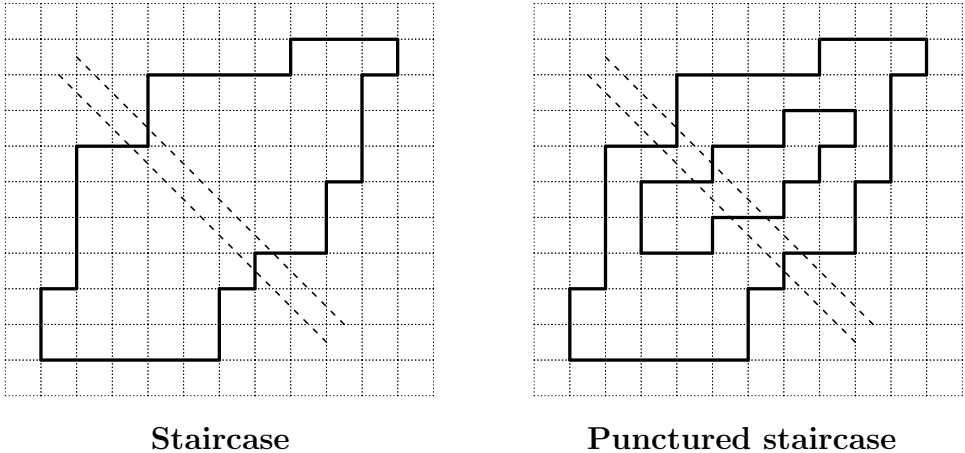}
\end{center}
\caption{\label{fig:poly} Examples of the types of polygons studied in this paper.
}
\end{figure}

A staircase polygon can be viewed as the intersection of two directed walks starting at the
origin, moving only to the right or up and terminating once the walks join at a vertex. 
It is well-known that the generating function for staircase polygons is
$$P(x) = \frac{1-2x-\sqrt{1-4x}}{2} \propto (1-\mu x)^{2-\alpha},$$
where the connective constant $\mu=4$ and the critical exponent $\alpha=3/2$.
Punctured staircase polygons \cite{GJWE00} are staircase polygons with internal holes which are 
also staircase polygons (the polygons are mutually- as well as  self-avoiding). In \cite{GJWE00} 
it was proved that the connective constant $\mu$ of $k$-punctured 
polygons (polygons with $k$ holes) is the same as the connective constant of unpunctured polygons. 
Numerical evidence clearly indicated that the critical exponent $\alpha$ increased by $3/2$ per 
puncture. The closely related model of punctured discs was considered in \cite{JvRW90}.
Punctured discs are counted by area and in this case it was proved that 
the critical exponent increases by 1 per puncture.
Here we will study only the case with a {\em single} hole (see figure~\ref{fig:poly}),
and we will refer to these objects as punctured staircase polygons. The perimeter length of
staircase polygons is even and thus the total perimeter (the outer perimeter plus the perimeter
of the hole) is also even. We denote by $p_n$ the number of punctured staircase polygons 
of perimeter $2n$. The results of \cite{GJWE00} imply that the half-perimeter generating 
function has a simple pole at $x=x_c=1/\mu=1/4$, though the analysis in \cite{GJWE00}
clearly indicated that the critical behaviour is more complicated than a simple algebraic
singularity.

Recently we found that the perimeter generating function of three-choice polygons
can be expressed as the solution of an 8th order linear ODE \cite{GJ06a}.
Similarly, in this paper we report on work which has led to an exact Fuchsian linear
differential equation of order 8 apparently satisfied by the perimeter generating function, 
$\PP(x) = \sum_{n\geq 0} p_nx^n$, for punctured staircase polygons (that is, $\PP (x)$ is
one of the solutions of the ODE, expanded around the origin). The first few terms in the
generating function are 
$$\PP(x) = x^8 + 12x^9+94x^{10}+604x^{11}+3463x^{12}+\cdots.$$
Our analysis of the ODE shows that the dominant singular behaviour is
\begin{equation}
\PP(x) \sim \frac{A(x)}{(1-4x)}  + \frac{B(x) + C(x) \log(1-4x)}{\sqrt{1-4x}}+D(x) (1+4x)^{13/2}.
\end{equation}
So in the notation used above, the generating function has a dominant singularity
at $x=x_c=1/\mu$ with exponent $\alpha=3$. 
This result confirms exactly  the conjecture  for the critical exponent \cite{GJWE00} 
in the case of a single puncture and the quite complicated corrections at the critical point
explains why the analysis in \cite{GJWE00}, based on a relatively short series, was so difficult.

\begin{figure}
\begin{center}
\includegraphics[scale=0.5]{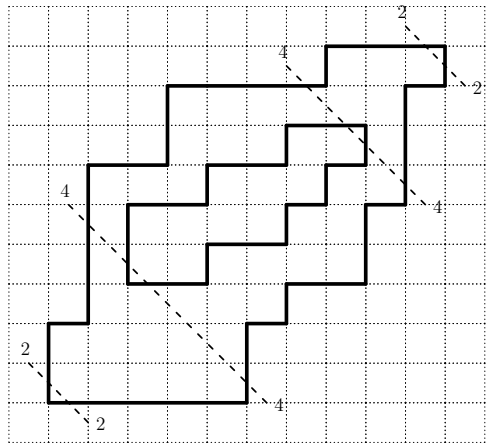}
\end{center}
\caption{ \label{fig:gv} The decomposition of a punctured staircase polygon
into a sequence of 2-4-2 vicious walkers, each expressible as a Gessel-Viennot determinant}
\end{figure}

It is also possible to express the generating function $\PP(x)$ as a sum of $4 \times 4$ 
Gessel-Viennot determinants \cite{GV89}. This is clear from figure \ref{fig:gv}, 
where the enumeration of the lattice paths between the dotted lines is just
the classical problem of 4 vicious walkers, and these must be joined to two
vicious walkers to the left, and to two vicious walkers to the right. Then one must sum over 
different possible geometries. However the fact that the generating function is so expressible 
implies that it is differentiably finite \cite{Lipshitz89}. 

Unfortunately we cannot readily bound the size of the underlying ODE, otherwise we could
use this observation to provide a proof of our results. As it is, we originally generated the counts 
of punctured staircase polygons up to perimeter 502 (251 coefficients), and found what we 
believe to be the underlying ODE experimentally from the first 195 coefficients. The ODE 
then correctly predicts the next 56 coefficients. While the possibility that the underlying 
ODE is not the correct one is extraordinarily small, our procedure still does not constitute 
a proof of course. We have since extended the count beyond perimeter 800 and still all 
coefficients are predicted by our ODE.

\section{Computer enumeration \label{sec:enum}}

The algorithm we use to count the number punctured staircase 
polygons is a modified version of the algorithm of Conway \etal \cite{CGD97}
for the enumeration of imperfect staircase polygons.
The two problems are very similar and consequently there are only minor differences
between the algorithms. A detailed description of the algorithm we used to
count imperfect staircase polygons can be found in \cite{GJ06a}.
The algorithm is based on transfer matrix techniques. This entails
bisecting the polygons by a line (as illustrated in figure~\ref{fig:poly}) and
enumerating the number of polygons by moving the line `forward' one step
at a time. Punctured staircase polygons start out as ordinary staircase polygons
and the line bisects the polygon at two edges. Then at some vertex two additional 
directed walks (sharing the same starting point) are inserted between the two 
original walks. The line will thus intersect these polygon configurations at 
four edges (see figure~\ref{fig:poly}).  The only difference between the algorithm 
in \cite{GJ06a} and the one used for this paper is in how the four directed walks  
intersected by the line are connected in order to produce a valid polygon. 
To produce a punctured staircase polygon we first connect the two 
innermost walks and then the two outermost walks are connected.   Imperfect staircase
polygons on the other hand are produced by connecting the first walk with the second walk 
and the third walks with the fourth walk. The updating rules used to count 
imperfect staircase polygons are given in \cite{GJ06a} and are easily amended to count
punctured staircase polygons bearing in mind the different `closing' criteria outlined
above.

We calculated the number of punctured staircase polygons up to perimeter 502.  
The integer coefficients become very large so the calculation was performed using 
modular arithmetic \cite{KnuthACPv2}. This involves performing the calculation {\it modulo} 
various prime numbers $p_i$ and then reconstructing the full integer coefficients 
at the end. We used primes of the form $p_i=2^{30}-r_i$, where $r_i$ are small positive
integers, less than $1000,$ chosen so that $p_i$ is prime, and $p_i \ne p_j$ unless $i = j.$
17 primes were needed to represent the coefficients correctly. The calculation for each 
prime used about 200Mb of memory and about 8 minutes of CPU time on a 
2.8 GHz Xeon processor. Naturally we could have carried the calculation much further
(and we have since done this) but as we shall demonstrate in the next section this number
of coefficients more than sufficed to identify an exact differential equation satisfied by $\PP (x)$.

\section{The Fuchsian differential equations \label{sec:fde}}

In recent papers Zenine {\it et al} \cite{ZBHM04a,ZBHM05a,ZBHM05b} obtained the
linear differential equations whose solutions give the 3- and 4-particle contributions
$\chi^{(3)}$ and $\chi^{(4)}$ to the Ising model susceptibility. In \cite{GJ06a} we
used their method to find a linear differential equation for three-choice polygons
and in this paper we extend this work further to find a linear differential equation 
which has as a solution the generating function $\PP (x)$ for punctured staircase polygons. 
We briefly outline the method here. Starting from a (long) series 
expansion for the function $\PP (x)$ we look for a linear differential equation of order $m$ 
of the form

\begin{equation}\label{eq:de}
\sum_{k=0}^m P_k(x) \frac{\rmd^k}{\rmd x^k}\PP (x) = 0,
\end{equation}
such that $\PP (x)$ is a solution to this homogeneous linear differential equation, where
the $P_k(x)$ are polynomials. In order to make it as simple as possible we start
by searching for a Fuchsian  \cite{Ince} equation. Such equations have only regular singular points. 
There are several reasons for searching for a Fuchsian equation, rather than a more general D-finite
equation. Computationally the Fuchsian assumption simplifies the search for a  solution.
>From the general theory of Fuchsian  \cite{Ince} equations it follows 
that the degree of $P_k(x)$ is at most $n-m+k$ where $n$ is the degree of $P_m(x)$. 
To simplify matters further (reduce the order of the unknown polynomials) it is advantageous
to explicitly assume that the origin and $x=x_c=1/4$ are regular singular points
and set $P_k(x)=Q_k(x)S(x)^k$, where $S(x)=x(1-4x)$. Thus when searching for a solution 
of Fuchsian type there are only two parameters, namely the order $m$ of the differential 
equation and the degree $q_m$ of the polynomial $Q_m(x)$.
One may also argue, less precisely, that for ``sensible''  combinatorial models one
would expect Fuchsian equations, as irregular singular points are characterized by explosive,
super-exponential behaviour. Such behaviour is not normally characteristic of combinatorial
problems arising from statistical mechanics. The point at infinity may be an exception to 
this somewhat imprecise observation.

We then search systematically for solutions by varying $m$ and $q_m$. In this
way we first found a solution with $m=10$ and $q_m=11$, which required the determination
of $L=195$ unknown coefficients. We have 251 terms in the half-perimeter series and
thus have 56 additional terms with which to check the correctness of our solution.
Having found this solution we then turned the ODE into a recurrence relation and used this
to generate more series terms in order to search for a lower order Fuchsian equation.
The lowest order equation we found was eighth order and with $q_m=27$, which requires the determination
of $L=294$ unknown coefficients. Thus from our original 251 term series we could not have found
this $8^{th}$ order solution since we did not have enough terms to determine all the unknown coefficients
in the ODE. This raises the question as to whether perhaps there is an ODE of lower order than 8 that
generates the coefficients? The short answer to this is no. Further study of our differential 
operator revealed that it can be factorised. In fact we found a factorization into three first-order
linear operators, a second order and a third order. The generating function is a solution of
the $8^{th}$ order operator, not of any of the smaller factors.

The (half)-perimeter generating function $\PP (x)$ for punctured staircase polygons
is a solution to the linear differential equation of order 8

\begin{equation}
\sum_{k=0}^8 P_n(x) \frac{\rmd^k}{\rmd x^k}\PP (x) = 0
\label{eq:PPfde}
\end{equation}
with

\begin{eqnarray}
P_8(x)=x^4(1-4x)^8(1 + 4x)(1 + 4x^2)(1 + x + 7x^2)Q_8(x). \nonumber \\
P_7(x)=x^3(1-4x)^7 Q_7(x), \;\;\;\;\;\; P_6(x)=2x^2 (1-4x)^6 Q_6(x), \nonumber \\
P_5(x)=6x(1-4x)^5 Q_5(x), \,\;\;\;\;\; P_4(x)=120(1-4x)^4 Q_4(x), \label{eq:PPpol} \\
P_3(x)=120(1-4x)^3 Q_3(x), \;\;\;\; P_2(x)=720(1-4x)^2 Q_2(x), \nonumber \\
P_1(x)=720(1-4x) Q_1(x), \;\;\;\;\; P_0(x)=2880 Q_0(x), \nonumber
\end{eqnarray}
where $Q_8(x)$, $Q_7(x)$, $\ldots$, $Q_0(x)$, are polynomials of degree 22, 28, 29, 30, 31,
31, 31, 31, and 31, respectively.  The polynomials are listed in \ref{app:PPpol}.

The singular points of the differential equation are given by the roots of $P_8(x)$.
One can easily check that all the singularities (including $x=\infty$) are
{\em regular singular points} so equation (\ref{eq:PPfde}) is indeed of the Fuchsian type.
It is thus possible using the method of Frobenius to obtain from the indicial equation
the critical exponents at the singular points. These are listed in Table~\ref{tab:PPexp}.

\Table{\label{tab:PPexp}
Critical exponents for the regular singular points of the Fuchsian differential
equation satisfied by $\PP (x)$.}
\br
Singularity & Exponents \\
\mr
$x=0$ & $-1, \, 0, \, 0, \, 0, \, 1, \, 2, \, 3, \, 8$ \\
$x=1/4$ & $-1$, \, $-1/2, \, -1/2, \, 1/2, \, 1, \, 3/2, \, 2, \, 3$ \\
$x=-1/4$ & $0, \, 1, \, 2, \, 3, \, 4, \, 5, \, 6, \, 13/2$ \\
$x=\pm\, \rmi/2$ & $0, \, 1, \, 2, \, 3, \, 4, \, 5, \, 6, \, 13/2$ \\
$1+x+7x^2=0$ & $0, \, 1, \, 2, \, 2, \, 3, \, 4, \, 5, \, 6$ \\
$1/x=0$ & $-2, \, -3/2, \, -1, \, -1, \, -1/2, \, 1/2, \, 3/2, \, 5/2$ \\
$Q_8(x)=0$ & $0, \, 1, \, 2, \, 3, \, 4, \, 5, \, 6, \, 8$ \\
\br
\endTable

We shall now consider the local solutions to the differential equation around each singularity. 
Recall that in general it is known \cite{ForsythV4,Ince} that if the indicial equation yields $k$ 
critical exponents which differ by an integer, then the local solutions {\em may} 
contain logarithmic terms up to $\log^{k-1}$. However, for the Fuchsian equation (\ref{eq:PPfde})
{\em only} multiple roots of the indicial equation give rise to logarithmic terms in the local 
solution around a given singularity, so that a root of multiplicity $k$  gives rise to
logarithmic terms up to $\log^{k-1}$. 
In particular this means that near any of the 22 roots of $Q_8(x)$ the local solutions
have no logarithmic terms and the solutions are thus {\em analytic} since all the
exponents are positive integers. The roots of $Q_8$ are thus {\em apparent singularities}
\cite{ForsythV4,Ince} of the Fuchsian equation (\ref{eq:PPfde}). We briefly
mention that as in our earlier study \cite{GJ06a} we can find a
solution of order 14 of the same form as (\ref{eq:PPfde}) but with $Q_{14}(x)$ being 
just a constant. So at this order none of the roots of $Q_8(x)$ appear. Clearly any real 
singularity of the system cannot be made to vanish and so we conclude that the  
22 roots of $Q_8$ must indeed be apparent singularities.

Assuming that only repeated roots give rise to $\log$-terms, and thus that a sequence
of positive integers give rise to {\em analytic} terms, then
near the physical critical point $x=x_c=1/4=1/\mu$ we expect the singular behaviour
\begin{equation}\label{eq:xc}
\PP(x) \sim  \frac{A(x)}{(1-4x)}  + \frac{B(x) + C(x) \log(1-4x)}{\sqrt{1-4x}},
\end{equation}
where $A(x)$, $B(x)$ and $C(x)$ are analytic in a neighbourhood of $x_c$. Note that the terms
associated with the exponents $1/2$ and $3/2$ become part of the analytic correction to the
$(1-4x)^{-1/2}$ term.
Near the singularity on the negative $x$-axis, $x=x_-=-1/4$ we expect the singular behaviour
\begin{equation}\label{eq:xm}
\PP(x) \sim D(x) (1+4x)^{13/2},
\end{equation}
where again $D(x)$ is analytic near $x_-$. We expect similar behaviour near the pair of 
singularities $x=\pm \rmi/2$, and finally at the
roots of $1+x+7x^2$ we expect the behaviour $E(x)(1+x+7x^2)^2\log (1+x+7x^2)$. 

We can simplify the $8^{th}$ order differential operator found above. We first found three
very simple solutions of the ODE, each corresponding to an order one differential operator,
$$
F_1(x)=1-4x,
$$
$$
F_2(x)=\frac{1-4x-4x^3}{1-4x},
$$
and
$$
F_3(x)=\frac{9-34x+14x^2}{\sqrt{1-4x}}.
$$
The existense of these three linearly independent solutions implies that we can find
three first order operators, which we denote by $L_i^{(1)},$ with $i =$ 1,2,3, such that 
the original 8'th order differential operator can be written as 
$L^{(8)}=L^{(5)}L_1^{(1)}L_2^{(1)}L_3^{(1)}$, where $L^{(5)}$ is a fifth order
differential operator.  We further found that this latter operator is decomposable as 
$L^{(5)}=L^{(3)}L^{(2)}$. This then allows one, in principle, to write
down the form of the $8 \times 8$ matrix representing the differential Galois group of
$L^{(8)}$, in an appropriate global solution basis.
To determine the asymptotics one would need to calculate non-local connection matrices
between solutions at  different points. 
This is a huge task for such a large differential operator.
Instead, we have developed a numerical technique that avoids all these difficulties,
which we describe below.

To standardise our asymptotic analysis, we assume that the critical point is at 1. The
growth constant of punctured staircase staircase polygons is 4, so we normalise
the series by considering the new series with coefficients $r_n$, defined by 
$r_n = p_{n+8}/4^n.$
Thus the generating function we study is
$\RP(y) = \sum_{n\geq 0} r_ny^n = 1 + 3y + 5.875y^2 + \cdots$.
Using the recurrence relations for $p_n$ (derived from the ODE) it is easy and fast
to generate many more terms $r_n$. We generated the first 100000 terms and saved 
them as floats with 500 digit accuracy (this calculation took less than 15 minutes).
>From equations (\ref{eq:xc}) and (\ref{eq:xm}) 
it follows that the asymptotic form of the coefficients is
\begin{equation}\label{eq:asymp}
[y^n]\RP(y) = r_n = \sum_{i \ge 0} \left( \frac{\tilde{a}_i}{n^i}\!+
\!\frac{\tilde{b}_i\log{n} \!+\! \tilde{c}_i}{n^{i+1/2}}
\!+\! (-1)^n\left( \frac{\tilde{d}_i}{n^{15/2+i}} \right)\! \right ) + {\rm O}(\lambda^{-n}).
\end{equation}
Any contributions from the other singularities are exponentially suppressed since
their norm (in the scaled variable $y=x/4$) exceeds 1.

Estimates for the amplitudes were obtained by fitting $r_n$ to the form given above using
an increasing number of amplitudes. `Experimentally' we find we need about the same total number of 
terms at $x_c$ and $-x_c$. So in the fits we used the terms with amplitudes 
$\tilde{a}_i$, $\tilde{b}_i$, and $\tilde{c}_i$, 
$i=0,\ldots,K$ and $\tilde{d}_i$, $i=0,\ldots,3K$. Going only to $K$ with the 
$\tilde{d}_i$ amplitudes results in much 
poorer convergence and going beyond $3K$ leads to no improvement.
For a given $K$ we thus have to estimate $6K+4$ unknown amplitudes. So we use the
last $6K+4$ terms $r_n$ with $n$ ranging from 100000 to $100000-6K-3$ and solve the
resulting $6K+4$ system of linear equations. We can also add extra terms to the asymptotic form and
check what happens to the amplitudes of the new terms. If these amplitudes are
very small it is highly likely that the terms are not truly present (if the
calculation could be done exactly these amplitudes would be zero). 

Doing this we found that all the amplitudes $\tilde{a}_i$ appear to be zero 
except that $\tilde{a}_0=1024$,
e.g., with $K=20$ we find that the estimates for the amplitudes 
$\tilde{a}_1<10^{-70}$, $\tilde{a}_2<10^{-60}$, 
$\tilde{a}_3<10^{-50}$, etc. So in all likelihood the amplitudes 
$\tilde{a}_i=0$ for $i>0$. This then leads us to
the refined asymptotic form
\begin{equation}\label{eq:asymptrue}
\fl
[y^n]\RP(y) = r_n = 1024\left( 1+ \frac{1}{\sqrt{n}}\sum_{i \ge 0} \left( \frac{b_i\log{n} + c_i}{n^{i}}
+ (-1)^n\left( \frac{d_i}{n^{7+i}} \right) \right ) \right ) + {\rm O}(\lambda^{-n}).
\end{equation}
In fits to this form we then used the terms with amplitudes $b_i$, and $c_i$, $i=0,\ldots,K$ and 
$d_i$, $i=0,\ldots,2K$. For a given $K$ we thus have to estimate $4K+3$ unknown amplitudes.
We find that the amplitude estimates are fairly accurate up to around $2K/3$.
We observed this by doing the calculation with $K=30$ and $K=40$ and then looking
at the difference in the amplitude estimates. For $b_0$ and $c_0$ the
difference is less than $10^{-120}$, while for $d_0$ the difference is less than $10^{-116}$.
Each time we increase the amplitude index by 1 we lose around six significant digits in accuracy.
With $i=18$ the differences are respectively around $10^{-14}$ and $10^{-11}$.

>From our very long series it is possible to obtain accurate numerical estimates of many of the 
amplitudes $b_i$, $c_i$, and $d_i$,  with a precision of more than 100 digits for the dominant
amplitudes, shrinking to around 10 digits for the the case when $i = 18$ (actually we could
probably have pushed this further but there would be little point). In this way 
we found that $b_0 = -\frac{6\sqrt{3}}{\pi^{3/2}}$, $b_1=\frac{305}{4\sqrt{3}\pi^{3/2}}$,
$b_2=\frac{86123}{192\sqrt{3}\pi^{3/2}}$,
$c_0 = 1.55210340048879105374\ldots$ and 
$d_0 = \frac{48}{\pi^{3/2}}$,$d_1 = -\frac{2610}{\pi^{3/2}}$,
$d_2 = \frac{640815}{8\pi^{3/2}}$, $d_3 = -\frac{116785575}{64\pi^{3/2}}$, 
$d_4 = \frac{70325480841}{2048\pi^{3/2}}$,
though we have not been able to identify $c_0$. 
These amplitudes are known to at least 100 digits accuracy.

The excellent convergence is solid evidence (though naturally not a proof) that
the assumptions leading to equation~(\ref{eq:asymp}) are correct. Further evidence
was obtained as follows:  We have already argued that the terms of the form $1/n^i$, $i>0$
are absent. We found similar results if we added terms like  $\log{n}/n^i$,
$\log^2{n}/n^{i/2}$ or additional
$\log{n}$ terms at $y=-1$. So this fitting procedure provides convincing
evidence that the asymptotic form (\ref{eq:asymptrue}), and thus the assumptions
leading to this formula, are correct.

\section{Conclusion and Outlook}

We have developed an improved algorithm for enumerating punctured staircase polygons. The
extended series, coupled with a search program that assumes the solution is a {\em Fuchsian}
ODE, enabled us to discover the underlying ODE, which was of $10^{th}$ order. We
did this without using 56 of the coefficients that we had generated. That
is to say, 56 of the known coefficients were unused, and so their values provided
a check on the solution. This
leads us to believe that we have found the correct ODE, as it reproduces the known, unused
coefficients. Subsequently we have extended this check to more than 200 unused coefficients.
Further refinement allowed us to find an $8^{th}$ order ODE. 

A numerical technique we have developed specifically for such problems then allowed us to 
find accurate numerical estimates for the amplitudes of the first several terms in the
asymptotic form of the coefficients, and then to conjecture their exact value.

We have also initiated an investigation of the {\em area} generating function. We expect this
to involve $q$-series, and thus far our investigations only lead us to believe that the area
generating function $A(q)$ is of the form
$$A(q) = (G(q) + H(q)\sqrt{1 - q/\eta})/[J_0(1,1,q)^2],$$
where $J_0(x,y,q)$ is a $q$-generalisation of the Bessel function, and occurs, for
example, in the solution of the problem of staircase polygons enumerated by perimeter \cite{BM96a}.
Here $q=\eta$ is the first zero of $J_0(1,1,q)$, and $G$ and $H$ are regular in the
neighbourhood of $q = \eta.$ The coefficients thus behave asymptotically as 
$$a_n = [q^n]A(q) \sim const. \eta^{-n}n.$$
In a subsequent publication we propose to investigate the area generating function more
fully, and hopefully obtain more insight into the properties of the ODE we have found for
the perimeter generating function.

Furthermore in work with C. Richard \cite{RJG06} we have conjectured the scaling function for 
punctured polygons with an arbitrary number of punctures. We briefly review the properties of 
the two-variable area-perimeter generating function for staircase polygons.  
Of special interest is the point { $(x_c,1)$} where two lines of singularities meet.
The behaviour of the singular part of the generating
function about  { $(x_c,1)$} is expected to take the special form
$$
{ P(x,q) \sim P^{(reg)}(x,q) + (1-q)^\theta F((x_c-x)(1-q)^{-\phi}), \qquad (x,q) \nearrow,}
$$
where { $F(s)$} is a {\it scaling function} of combined argument { $s=(x_c-x)(1-q)^{-\phi}$,}
commonly assumed to be regular at the origin,
and  { $\theta = 1/3$} and { $\phi = 2/3$} are {\it critical exponents}.
For staircase polygons, we have
$$
{ F(s) = \frac{1}{8}\frac{\rmd}{\rmd s}\log\mbox{Ai} \left(  (4\sqrt{2})^\frac{2}{3} s \right)}.
$$
In  \cite{RJG06} we assumed that the limit distribution by area of staircase polygons 
with $r$ punctures (of arbitrary size) is that of staircase polygons with $r$ holes 
of unit area. From this and knowledge of $F(s)$ we then obtained {\em exact} 
predictions for $r$ punctured staircase polygons by taking  the $r$-th derivative w.r.t $q$ 
of $P(x,q)$. We then study the area-moment generating function,
$P_k (x) = \sum_{m,n} n^k p_{m,n}x^m$, where $p_{m,n}$ is the number of polygons
with perimeter $m$ and area $n$.  
In particular we find that the leading amplitudes $A^{(r)}_{k}$ of the perimeter 
generating function of the $k$-th area-moment are given by 
$$
A^{(r)}_{k}=\frac{(k+r)! x_c^r f_{k+r}}{r! x_c^{\gamma_{k+r}}\Gamma(\gamma_{k+r})}
$$
Here $f_{k+r}$ are amplitudes occuring in the asymptotic expansion of 
$P(x,q)$ (these are known exactly for punctured staircase polygons) 
while $\gamma_{k+r}=3(k+r)/2-1/2$ are the critical exponents of the 
$k$th area-moment of $r$ punctured polygons.
These predictions have been confirmed for once punctured staircase polygons to a very high level 
of accuracy for moments up to $k=10$. The numerical analysis of the  area-moments relied
crucially on our knowledge of the singularity structure of the  perimeter generating function
as detailed in this paper.

\section*{E-mail or WWW retrieval of series}

The series for the generating functions studied in this paper 
can be obtained via e-mail by sending a request to 
I.Jensen@ms.unimelb.edu.au or via the world wide web on the URL
http://www.ms.unimelb.edu.au/\~{ }iwan/ by following the instructions.

\ack

We would like to thank N Zenine and J-M Maillard for access to their Mathematica
routines for identifying  differential equations and useful advice about their use.
We gratefully acknowledge financial support from the Australian Research Council.

\appendix

\section{\label{app:PPpol} Polynomials $Q_n(x)$ for punctured staircase polygons}

\begin{eqnarray*}
\fl Q_8(x) &=& -90720 + 1255590 x - 9538200 x^2 + 20394828 x^3 - 79106610 x^4  \\
\fl && + 1223958687 x^5 - 2943232056 x^6 + 17470357067 x^7 - 189472079743 x^8  \\
\fl && + 579172715823 x^9 - 1746461498616 x^{10} + 8400325324610 x^{11} \\
\fl && - 1591154327260 x^{12} - 111431714394808 x^{13} + 315517552430480 x^{14} \\
\fl && - 106489387477312 x^{15} - 938487878760384 x^{16} + 1628517397980288 x^{17} \\
\fl && - 2394531569420032 x^{18} + 2966185168205312  x^{19} \\
\fl && - 170238270849024 x^{20} - 699187344629760 x^{21} + 295462090506240 x^{22}
\end{eqnarray*}

\begin{eqnarray*}
\fl Q_7(x) &=& -1360800 + 23565780 x - 167569290 x^2 + 478254996 x^3 + 641052858 x^4 \\
\fl &&   + 14810951034 x^5 - 47034372339 x^6 - 166933659974 x^7 - 2552936187594 x^8 \\
\fl &&   + 6447911404224 x^9 + 14253364474478 x^{10} + 86598771199392 x^{11} \\  
\fl &&   + 362131239586500 x^{12} - 3860712252484892 x^{13} + 8993313236994576 x^{14} \\
\fl &&   - 31235880957264960 x^{15} + 46429326957124912 x^{16} \\
\fl &&   + 155905775680790304 x^{17} - 807736441103822976 x^{18} \\ 
\fl &&   + 1835072857042276096 x^{19} - 1278888252797142528 x^{20} \\ 
\fl &&   - 293981468599460352 x^{21} +  14541716059525437440 x^{22} \\
\fl &&   - 26481815895022608384 x^{23} + 22483566008412450816 x^{24} \\
\fl &&   - 35911819535956066304 x^{25} + 3639680241277796352 x^{26} \\
\fl &&   + 7495959535363031040 x^{27} - 3507725938490081280 x^{28}
\end{eqnarray*}

\begin{eqnarray*}
\fl Q_6(x) &=& -1723680 + 69281730 x - 787195710 x^2 + 4886678970 x^3 - 10726639974 x^4 \\ 
\fl &&    + 11830409583 x^5 - 401281487235 x^6 + 343905413598 x^7 \\
\fl &&    + 1858137414650 x^8 + 44092692217413 x^9 - 36740412036168 x^{10} \\
\fl &&    - 135298590380414 x^{11} - 1279093006602396 x^{12} - 10004750418032976 x^{13} \\ 
\fl &&    + 61536871579988144 x^{14}  -216281351081049504 x^{15} \\
\fl &&    + 1050287576547538488 x^{16} - 1795967175346626976 x^{17} \\
\fl &&    - 2572736181692580960 x^{18} + 18017037664470796032 x^{19} \\
\fl &&    - 45232775265352713472 x^{20} + 48709527110201501184 x^{21} \\
\fl &&    + 4770083118869915136 x^{22} - 322327838255331590144 x^{23} \\
\fl &&    + 541571044899035842560 x^{24} - 511926023257614434304 x^{25} \\
\fl &&    + 716375351150156644352 x^{26} - 69659801950830723072 x^{27} \\
\fl &&    - 136551990333116252160 x^{28} + 60094625512245166080 x^{29}
\end{eqnarray*}

\begin{eqnarray*}
\fl Q_5(x) &=& 1965600 - 6539400 x - 358033410 x^2 + 4831433820 x^3 - 30915098190 x^4 \\
\fl &&   + 60211846008 x^5 - 201764518161 x^6 + 2531858233470 x^7 \\
\fl &&   + 1380416576424 x^8 - 20212314275250 x^9 - 61506470769366 x^{10} \\
\fl &&   - 477804842150324 x^{11} + 608746761166938 x^{12} + 483723642457152 x^{13} \\
\fl &&   + 60127368616743592 x^{14} - 185780400624937008 x^{15} \\
\fl &&   + 1165835175099337288 x^{16} - 7175943616536571776 x^{17} \\
\fl &&   + 13745698284061066112 x^{18} + 4948349174336379840 x^{19} \\ 
\fl &&   - 89453290124304769024 x^{20} + 270104157697832561664 x^{21} \\
\fl &&   - 356324521463829808128 x^{22} - 41862184650482117632 x^{23} \\
\fl &&   + 1845216328946812827648 x^{24} - 2906213125616330383360 x^{25} \\ 
\fl &&   + 2943265956913569742848 x^{26} - 3723507915329643413504 x^{27} \\
\fl &&   + 405249143061461336064 x^{28} + 618215144006850969600 x^{29} \\
\fl &&   - 261821958729561538560 x^{30}
\end{eqnarray*}

\begin{eqnarray*}
\fl Q_4(x) &=& 241920 - 8017380 x + 88351704 x^2 - 590355612 x^3 + 2409400818 x^4 \\
\fl &&   - 8457027588 x^5 + 71232186468 x^6 - 288557341128 x^7 \\
\fl &&   + 524905454055 x^8 - 5046532132734 x^9 + 28114089314043 x^{10} \\
\fl &&   - 164508486596467 x^{11} + 869331744354740 x^{12} - 2401501341116904 x^{13} \\
\fl &&   + 12275987679372578 x^{14} - 50846889626226508 x^{15} \\
\fl &&   + 46258831828476364 x^{16} - 147764159295056304 x^{17} \\
\fl &&   + 1375769527659995736 x^{18} - 2625251094439093408 x^{19} \\
\fl &&   - 765792895039661984 x^{20} + 22951686058011476032 x^{21} \\
\fl &&   - 85054223999548283904 x^{22} + 126294091912315062016 x^{23} \\
\fl &&   + 19381267403906712064 x^{24} - 566287434634380073984 x^{25} \\
\fl &&   + 849895463062111623168 x^{26} - 892557255237919469568 x^{27} \\
\fl &&   + 1043719341871898804224 x^{28} - 142670999896790335488 x^{29} \\
\fl &&   - 140350544778022354944 x^{30} + 59234239904690995200 x^{31}
\end{eqnarray*}

\begin{eqnarray*}
\fl Q_3(x) &=& -4596480 + 112443660 x - 1327020156 x^2 + 11580963786 x^3 - 76436209584x^4 \\
\fl &&   + 426159579924 x^5 - 2350462539072 x^6 + 11395385983233 x^7 \\
\fl &&   - 44136036344190 x^8 + 145288111685523 x^9 - 559910802106640 x^{10} \\
\fl &&   + 3013037795053530 x^{11} - 13499762948930634x^{12} \\
\fl &&   + 50096716464628528 x^{13} - 217987216302493908 x^{14} \\
\fl &&   + 853439326193439492 x^{15} - 2363497210984795232 x^{16} \\
\fl &&   + 5096223845046539304 x^{17} - 8508469151526998016 x^{18} \\
\fl &&   + 9581930085552894304 x^{19} - 10241374665198721536 x^{20} \\
\fl &&   - 12641088914996048640 x^{21} + 118651673978481267200 x^{22} \\
\fl &&   - 208768950136609496064 x^{23} - 15400291418459486208 x^{24} \\
\fl &&   + 814317146169694152704 x^{25} -1202858442211165741056 x^{26} \\
\fl &&   + 1271933402411862171648 x^{27} - 1406355411740766470144 x^{28} \\
\fl &&   + 251165051564655771648 x^{29} + 137326949251639934976 x^{30} \\
\fl &&   - 61285928661166325760 x^{31}
\end{eqnarray*}

\begin{eqnarray*}
\fl Q_2(x) &=& 1209600 - 10784340 x + 25225200 x^2 - 192390408 x^3 + 1497608946 x^4 \\
\fl &&   - 3085618896 x^5 + 55270573062 x^6 - 674664767886 x^7 + 1891951243653 x^8 \\
\fl &&   + 6937954472784 x^9 - 19443421819978 x^{10} - 252270853719194 x^{11} \\
\fl &&   + 1421753108033868 x^{12} - 2280488850916676 x^{13} - 1040351739238056x^{14} \\
\fl &&   - 1519080794794788 x^{15} + 54144924827952720 x^{16} \\
\fl &&   - 143110935850986376 x^{17} - 63031554528921744 x^{18} \\
\fl &&   + 1125126938486807936 x^{19} - 2675665192031509504 x^{20} \\
\fl &&   + 3361130538055156224 x^{21} - 2669659667713374208 x^{22} \\
\fl &&   + 1996890960732463104 x^{23} - 4866848788151009280 x^{24} \\
\fl &&   + 3555378162093901824 x^{25} + 3193922372633202688 x^{26} \\
\fl &&   - 2642707373157531648 x^{27} + 2132642211038560256 x^{28} \\
\fl &&   - 3311881541411143680 x^{29} + 1596569887904366592 x^{30} \\
\fl &&   - 264734033093591040 x^{31}
\end{eqnarray*}

\begin{eqnarray*}
\fl Q_1(x) &=& -725760 + 19969740 x - 254689092 x^2 + 2329185726 x^3 - 17948325636x^4 \\
\fl &&   + 118028863386 x^5 - 679983561900 x^6 + 3637871524611 x^7 \\
\fl &&   - 17150360490738 x^8 + 62088405193554 x^9 -183555964459890 x^{10} \\
\fl &&   + 747009873725220 x^{11} - 4106684548673028 x^{12} + 18540613780587884 x^{13} \\
\fl &&   - 67936944600058776 x^{14} + 247341581626824360 x^{15} \\
\fl &&   - 939866071520217104 x^{16} + 3216462341735279616 x^{17} \\
\fl &&   - 8789133587934808704 x^{18} + 17976423995943224576 x^{19} \\
\fl &&   - 26625353996773725696 x^{20} + 29354499014436664320 x^{21} \\
\fl &&   - 26197184327864145920 x^{22} + 20118012206750361600 x^{23} \\
\fl &&   - 11595016904008224768 x^{24} - 12803308242930466816 x^{25} \\
\fl &&   + 49275320633035751424 x^{26} - 49679788190366564352 x^{27} \\
\fl &&   + 31169615491025600512 x^{28} - 29010025645678264320 x^{29} \\
\fl &&   + 12772559103234932736 x^{30} - 2117872264748728320 x^{31}
\end{eqnarray*}

\begin{eqnarray*}
Q_0(x) = Q_1(x)
\end{eqnarray*}

%\bibliographystyle{ioptitle_unsrt}
%\bibliography{animals,saw,sap,spin,series}

\end{document}